# Spatial Information Refinement for Chroma Intra Prediction in Video Coding

Chengyi Zou[1], Shuai Wan[1], Tiannan Ji[1], Marta Mrak[2], Marc Gorriz Blanch[2], Luis Herranz[3]
[1]Northwestern Polytechnical University, Xi'an, China
E-mail: cyzou@mail.nwpu.edu.cn, swan@nwpu.edu.cn, jitiannan@mail.nwpu.edu.cn
[2]British Broadcasting Corporation, London, UK
E-mail: {marta.mrak, marc.gorrizblanch}@bbc.co.uk
[3]Computer Vision Center, Barcelona, Spain
E-mail: lherranz@cvc.uab.es

*Abstract*— Video compression benefits from advanced chroma intra prediction methods, such as the Cross-Component Linear Model (CCLM) which uses linear models to approximate the relationship between the luma and chroma components. Recently it has been proven that advanced cross-component prediction methods based on Neural Networks (NN) can bring additional coding gains. In this paper, spatial information refinement is proposed for improving NN-based chroma intra prediction. Specifically, the performance of chroma intra prediction can be improved by refined down-sampling or by incorporating location information. Experimental results show that the two proposed methods obtain 0.31%, 2.64%, 2.02% and 0.33%, 3.00%, 2.12% BD-rate reduction on Y, Cb and Cr components, respectively, under All-Intra configuration, when implemented in Versatile Video Coding (H.266/VVC) test model.

*Index Terms*-Chroma intra prediction, convolutional neural networks, spatial information refinement.

## I. INTRODUCTION

The latest video coding standard Versatile Video Coding (H.266/VVC) achieves about 50% bit-rate saving compared to its predecessor by introducing a number of new coding tools [1]. Cross Component Linear Model (CCLM) is one of the featured tools for intra chroma prediction. Prior to VVC, luma reconstruction values were not used to obtain chroma prediction in widely used compression profiles. In fact, there is a strong correlation between the luma and chroma components of the common video sequences, and that correlation can be further exploited to improve coding efficiency.

The purpose of CCLM is to improve the prediction accuracy of chroma components by exploiting cross-component correlation. A cross-component linear model is first built between the luma and chroma components, and then the chroma prediction is obtained by applying the linear model to the down-sampled luma reconstruction values, which contributes to a performance improvement of 1.54%, 13.89%, 14.76% on Y, Cb and Cr components, respectively, under All-Intra configuration [2]. A linear model can significantly improve coding efficiency at low complexity, however, it cannot always accurately describe the relationship between the luma and chroma components, and therefore cannot effectively predict coding units with complex textures.

Deep learning is useful for modeling the complex relationship between luma and chroma. Neural networks generally make the prediction in a data-driven manner. In [3], feature maps are generated by fusing boundary and current luma block information using an attention-based neural network, which guides the transfer of luma information through the network. However, the predictive ability of convolutional neural network (CNN) is limited, and simply deepening the network does not lead to more accurate predictions, while introducing significant complexity into the mode selection process at the encoder side. On the other hand, side information is useful for performance improvement. In this spirit, this paper improves chroma intra prediction by spatial information refinement, building upon the network of [3]. The contributions of this paper are summarized as follows.

- Considering different importance of boundary information for the predicted chroma pixel at each location, we propose a method for fusing reconstructed value data with the location information.
- Using a convolutional layer instead of original chroma down-sampling filter to better preserve features in luma information.

The remainder of the paper is organized as follows. A brief description of the related work about cross-component prediction is provided in Section II. In Section III, we describe the proposed spatial information refinement for chroma intra prediction. The performance of the proposed method is presented in Section IV. Finally, Section V concludes this paper.

## II. RELATED WORK

This section begins with a brief review of related work on cross-component linear model. Then, NN-based cross-component prediction is introduced.

### A. Cross-component Linear Model

The development of cross-component prediction began with research in high-resolution reconstruction and mosaic removal applications, in which substantial redundancy between the luma and chroma components was verified. In video coding, Lee and Cho [4] first found that the relationship between the luma and chroma components could be approximated by a linear model. In this algorithm, the linear model parameters were sent explicitly or derived implicitly. Based on H.265/HEVC, Khairat et al. [5] designed several enhanced luma-chroma prediction algorithms, called Cross Component Prediction (CCP). Due to the superior performance of the algorithms, their application in the residual domain has been adopted by the HEVC Range-Extension [6].

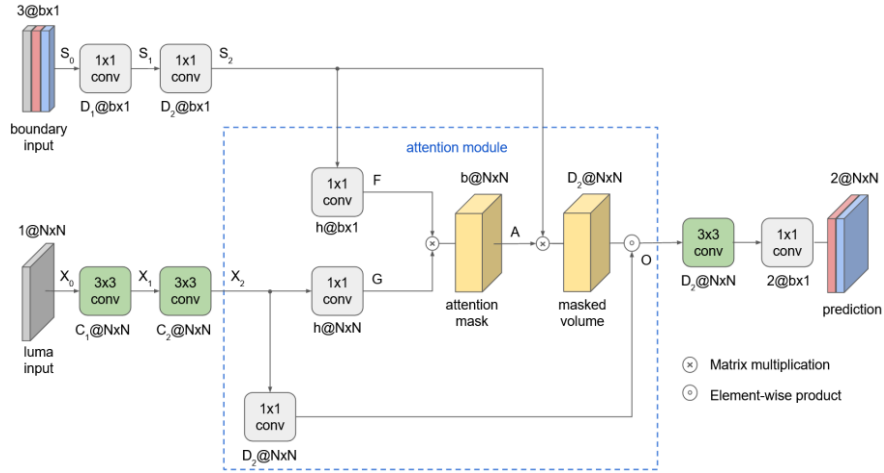

Fig. 1. Baseline attention-based architecture for chroma intra prediction presented in [3].

The outstanding performance of cross-component prediction has attracted the attention of many researchers. Gisquet and François [7] analyzed and solved the problem of inappropriate choice of reference samples in the derivation of linear models. Zhang et al. [8] proposed to use only left-adjacent or upper-adjacent sampling to derive linear models. Zhang et al. [9] proposed an adaptive template selection method to extend the adjacent sampling in the LM mode, where the Cr component can be predicted from either the Y component or the Cb component, or from both the Y and Cb components. Lee et al. [10] used the weighted sum of adjacent pixels as the prediction for the current chroma pixel, with the weighting factor determined by the relationship between luma and chroma. Yeo et al. [11] perform template matching using co-located luma block and then use the co-located chroma block of the matching luma block as a predictor of the current chroma block.

Pu et al. [12] proposed a new inter-component de-correlation method performed in the residual domain, using the luma residual signal to compensate for the chroma residual signal. Zhang et al. [13, 14] attempt to model the relationship between luma and chroma more accurately, proposing a method known as Multi Model CCLM (MMLM). The method starts with the derivation of $N$ sets of linear models by means of a linear regression algorithm. The reconstructed luma values of the current block is then divided into $N$ groups and a linear model is applied to each group. Secondly, more filters are designed for the down-sampling of luma. Thirdly, an LM-angular prediction method is proposed, which combines the angular intra prediction and the CCLM prediction.

### B. NN-based Cross-component Prediction

The relationship between luma and chroma cannot often be described by a linear model, and neural networks can be used to model a more complex nonlinear mapping. Non-linear activation function and iterative update of parameters provide the network with an excellent learning ability.

Li et al. [15] used fully connected layers to extract information about the boundary reference pixels and CNN layers to extract information about the down-sampled reconstructed luma. Both types of features were then fused to produce a prediction of the chroma. Meyer et al. [16] applied a different approach for data processing. Unlike the former, which treats the boundary and current block separately, the approach in [16] used pixels of luma and chroma as inputs to the two branches of the network. Zhu et al. [17] made use of more reference pixels for prediction at the Coding Tree Unit (CTU) level. The number of calculations was reduced by copying the predicted values at the corresponding position at the CTU level when predicting at the Coding Unit (CU) level. The CCLM predictions were used as the initial values and the network learnt the residuals between the predictions and the original values. QP information was introduced on the input side of the network. The pipeline of coding is, however, altered by this method, and more information need to be stored at both the encoding and decoding side.

An attention-based chroma prediction network was proposed by Blanch et al. [3, 18], which is more friendly to a standardized implementation. Fig. 1. shows the corresponding network structure. There are mainly 4 branches in this network, i.e., cross-component boundary branch, luma convolutional branch, attention-based fusion branch and prediction head branch. The first two branches extract cross-component information and luma space information. The third branch merges the two parts of features, and the final branch obtains the predicted values of Cb and Cr. The network can be trained with blocks of different sizes to adapt to the CU size.

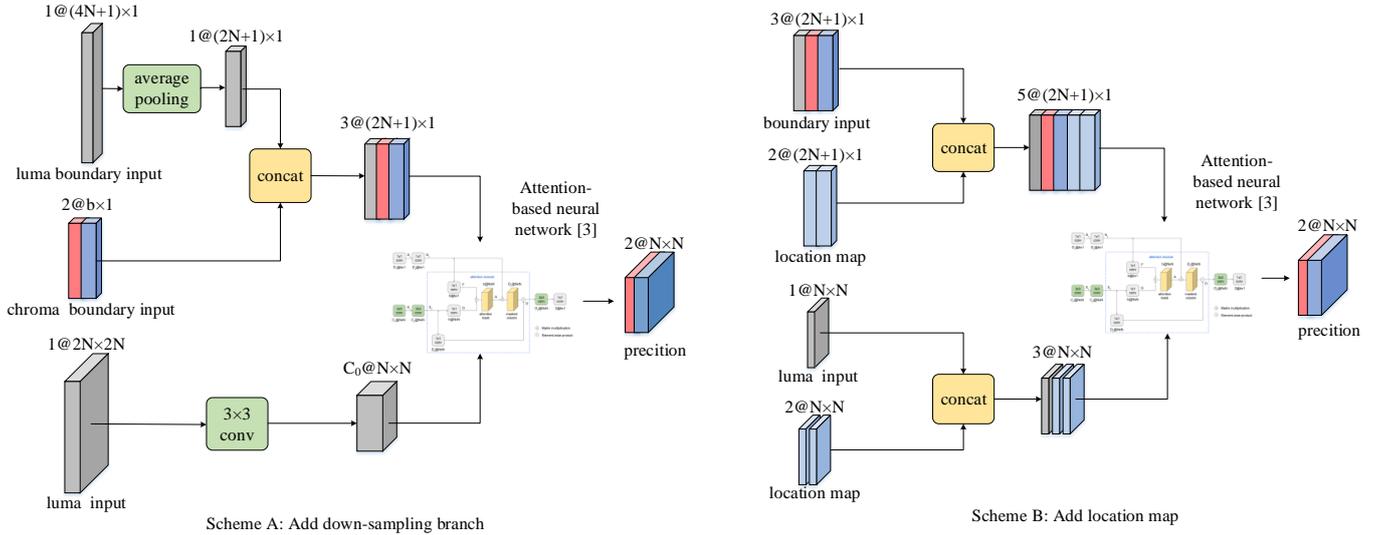

Fig. 2 Attention-based neural network with proposed enhanced chroma prediction methods.

This work has also contributed to the simplification of networks and the reduction of the computational complexity of the encoder. Specifically, two convolutional layers with $3 \times 3$ kernel size are combined into one convolutional layer with a $5 \times 5$ kernel size by removing the middle non-linear activation function. Using the idea of autoencoder, the number of feature channels is reduced in the middle layer of the network. In the process of embedding the network into the encoder, the operations of displacement and table look-up are used to replace the original multiplication and division operations.

## III. PROPOSED METHOD

Building upon the attention-based neural network in [3], two schemes for spatial information refinement are proposed to improve the chroma prediction performance. By adding down-sampling branches or location information to the input, the network performance is improved as more accurate prediction values can be obtained.

The proposed methods are illustrated in Fig. 2. Without changing the main structure, the two proposed schemes enhance the cross-component boundary branch and the luma convolutional branch. The proposed two schemes are:
- Scheme A: Adding down-sample branch.
- Scheme B: Adding location map.

### A. Down-sampling branch

Unlike the down-sampling filter in CCLM, here the down-sampling filter is obtained through learning, in order to select most the suitable down-sampling luma feature for chroma prediction.

In this discussion, we consider 4:2:0 YUV format. The size of the luma block is assumed to be $2N \times 2N$, which is twice the size of the chroma block in both the width and the height. For the chroma prediction process, the reference samples used include the collocated luma block $X \in \mathbb{R}^{2N \times 2N}$, the reference sample array $B_y \in \mathbb{R}^{b_y}$ at the top-left corner of the current block, where $b_y = 4N+1$, and the reference sample array $B_c \in \mathbb{R}^{b_c}$ at the top-left corner of the current block with $b_c = 2N+1$. Here $y$ refers to the $Y$ component, and $c = Cb$ or $Cr$ to refer the two chroma components, respectively. $B$ is built from the samples on the left boundary (starting from the bottommost sample), then corner points are added, and finally the top samples (starting from the leftmost sample) are added. If some reference samples are not available, these are filled with the DC value which is defined as:

$$Value_{DC} = 2^{bitdepth-1}, \quad (1)$$

where $bitdepth$ is 10.

In order to concatenate the luma features and chroma features for chroma prediction, we need to down-sample the luma input to the same size as the chroma input. The luma down-sampling filter of the cross-component boundary branch selects the average pooling with a size of 2, as:

$$B'_y(i) = \frac{B_y(2i) + B_y(2i+1)}{2}, \quad (2)$$

where $i = 0, 1, \ldots, 2N$ represents the pixel position, and $B'_y \in \mathbb{R}^{2N+1}$ represents the down-sampled luma boundary feature.

The down-sampled luma feature $B'_y$ is concatenated with the chroma boundary feature $B_{cb}$ and $B_{cr}$ to obtain the cross-component volume $S \in \mathbb{R}^{3 \times b}$, which is the input of the cross-component boundary branch. Note that $B_y$ fills a pixel to ensure the same size as $B_{cb}$ and $B_{cr}$ after the pooling operation.

The luma down-sampling filter of the luma convolutional branch uses $3 \times 3$ convolution with a step size of 2, which expands the channel of the feature map while down-sampling. Similar to subsequent consecutive $C_j$-dimensional $3 \times 3$ convolution, a bias and a Rectified Linear Unit (ReLU) non-linear activation function are applied as:

$$X_1 = \mathrm{Re}LU(W*X+b), \qquad (3)$$

where $W \in \mathbb{R}^{1 \times D_1}$ and $b$ are the weights and bias respectively, $X_1 \in \mathbb{R}^{N \times N \times D_1}$ represents the feature of luma current block after down-sampling and also the input of the luma convolutional branch.

### B. Location information fusion

In the prediction process, the accuracy of the prediction of pixels at different spatial position will be affected by the corresponding distance from the reference sample line(s) However, the calculation of convolution is the same for each convolution block, and spatial differences are not considered in [3]. In order to allow the network to predict the pixels at different positions of the current block at different importance levels, and to effectively guide the prediction process, we use the position information of the current block and the boundary to construct a feature map, called location map, as shown in the Fig. 3. The location map is the same as the input size, with values of normalized abscissa and ordinate information of each pixel.

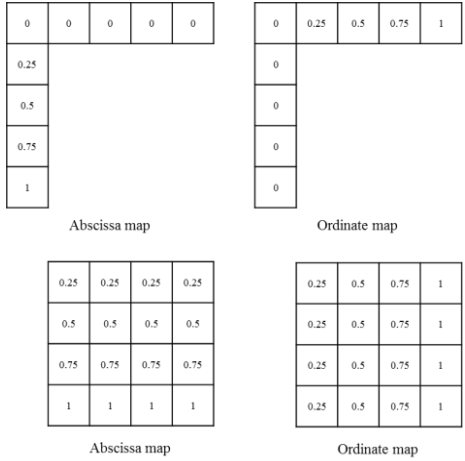

Fig. 3 Location information for a 4 by 4 block; the top is the location map of the boundary, and the bottom is the location map of the current block.

In this discussion, the size of the down-sampled luma block is assumed to be $N \times N$ which is the same size as the chroma block. For the chroma prediction process, the collocated luma block $X$ becomes $X \in \mathbb{R}^{N \times N}$, the reference sample array $B$ becomes $B \in \mathbb{R}^{b_c}$, $b_c = 2N+1$, $c$ refers to the three components of $Y$, $Cb$ and $Cr$ respectively, the input of the cross-component boundary branch is denoted by $S \in \mathbb{R}^{3 \times b}$, which is formed by concatenating the three reference arrays of $B_y$, $B_{cb}$, and $B_{cr}$. Before the NN processing, all the input features are merged with the location map:

$$\begin{aligned} S_1^{0-2} = S, S_1^3 = A_s, S_1^4 = O_s \\ X_1^0 = X, X_1^1 = A_x, X_1^2 = O_x \end{aligned}, \qquad (4)$$

where $A$ and $O$ are abscissa and ordinate map respectively, $S_1 \in \mathbb{R}^{5 \times b}$ and $X_1 \in \mathbb{R}^{N \times N \times 3}$ are the input of cross-component boundary branch and luma convolutional branch after adding location map, respectively.

It is worth mentioning that the two proposed methods are applied with a single line of reference pixels when samples in the double lines of reference samples are not available. On the one hand, this design reduces the amount of calculation, whether it is in down-sampling or subsequent prediction. On the other hand, using double boundary reference samples suffers from performance loss when double reference lines are unavailable, such as the top-right or bottom-left of the current block. If these reference samples are not available, the sample values will be filled with predefined values, which are most likely to be deviated from the available reference samples.

## IV. EXPERIMENT RESULTS

**Dataset setting:** Training examples were extracted from the DIV2K dataset [19], which contains high-resolution content with large diversity. The database contains 800 training samples and 100 samples for validation. It provides 6 lower-resolution versions, and uses bilinear and unknown filters to down-sample by factors 2, 3, and 4. For the method of adding down-sample branch, we convert the picture from PNG format to YUV4:2:0 format, and for the method of adding location map, we convert the picture from PNG format to YUV4:4:4 format. For each data instance, a resolution is randomly selected, and then blocks of each $N \times N$ size ($N = 4, 8, 16$) are selected, making balanced sets between the block size in each image and the uniform space selection. All the pixel values are normalized to [0,1].

**Training configuration:** We use Keras/Tensorflow to implement our model. During the training process, we set the batch size to 16. We use the Mean Square Error (MSE) between the output of the network and the ground truth image as the loss function, and the network of all proposed schemes uses Adam optimizer [20], and starts training with a learning rate of $10^{-4}$.

**Test and comparison:** We integrate the final models of two schemes into VTM7.0 respectively, joining the RD competition of chroma prediction as a new prediction mode, and use All-Intra configuration suggested by Common Test Conditions (CTC) [21]. QP is set to 22, 27, 32, 37. For a fair comparison with [3], only 4×4, 8×8 and 16×16 square blocks are supported. We use BD-rate to evaluate the relative compression efficiency. The test set includes 20 video sequences with different resolutions, called A, B, C, D, E category.

We summarize all the component-wise BD-rate results for the proposed schemes and the original attention-based approach in [3], as shown in Table I. Scheme A achieves an average Y/Cb/Cr BD-rates of -0.31%/-2.64%/-2.02% compared with the VTM-7.0 anchor, suggesting that the proposed down-sampling branch can improve the coding performance of the related attention-based model. As for the method that is based on location information fusion, as introduced in Scheme B, an average of -0.33%/-3.00%/-2.12% performance gain in Y/Cb/Cr BD-rates can be achieved compared with the VTM-7.0 anchor. We can observe that both schemes outperform, and in the proposed methods Scheme B achieves better performance.

TABLE I
THE RATE-DISTORTION PERFORMANCE COMPARISON IN AI CONFIGURATION. (ANCHOR: VTM-7.0)

| method | [3] | | | Scheme A | | | Scheme B | | |
|---|---|---|---|---|---|---|---|---|---|
| component | Y | U | V | Y | U | V | Y | U | V |
| Class A1 | -0.28% | -3.20% | -1.85% | -0.32% | -3.28% | -1.79% | -0.39% | -4.10% | -2.35% |
| Class A2 | -0.25% | -3.11% | -1.54% | -0.31% | -4.17% | -1.88% | -0.11% | -4.10% | -2.09% |
| Class B | -0.26% | -2.28% | -2.33% | -0.20% | -2.53% | -2.25% | -0.29% | -2.91% | -2.82% |
| Class C | -0.30% | -1.92% | -1.57% | -0.39% | -2.10% | -1.88% | -0.45% | -2.43% | -1.70% |
| Class D | -0.29% | -1.70% | -1.77% | -0.44% | -2.41% | -2.11% | -0.50% | -2.50% | -1.89% |
| Class E | -0.13% | -1.59% | -1.45% | -0.23% | -1.72% | -2.04% | -0.16% | -2.37% | -1.62% |
| Overall | -0.26% | -2.25% | -1.80% | -0.31% | -2.64% | -2.02% | -0.33% | -3.00% | -2.12% |

## V. CONCLUSIONS

This paper proposed two methods of spatial information refinement to enhance chroma prediction based on existing attention-based structure. The network can be improved by adding down-sampling branch or the location map. Experiments have proved that spatial information refinement is valid, our proposed methods achieve better performance. Future work will focus on spatial information refinement with reduced complexity.